\documentclass[reprint,aps,showpacs,prl]{revtex4-1}  

\usepackage{amsmath}
\usepackage{amssymb}
\usepackage{graphicx}
\usepackage{bm}
\usepackage{color}
\usepackage{times}

\newcommand{\etal}{\,\textit{et al.}}
\definecolor{Blue}{rgb}{0.08,0.17,0.55}

\begin{document}

\title{Cooperative optical non-linearity due to dipolar interactions in an ultra-cold Rydberg ensemble}

\author{J. D. Pritchard}
\email{j.d.pritchard@durham.ac.uk}
\author{A. Gauguet}
\author{K. J. Weatherill}
\author{M. P. A. Jones}
\author{C. S. Adams}
\email{c.s.adams@durham.ac.uk}
\affiliation{Department of Physics, Durham University, Rochester Building, South Road, Durham DH1 3LE, UK}

\date{\today}

\begin{abstract}
We demonstrate a cooperative optical non-linearity caused by dipolar interactions between Rydberg atoms in an ultra-cold atomic ensemble. By coupling a probe transition to the Rydberg state we map the strong dipole-dipole interactions between  Rydberg pairs onto the optical field. We characterize the non-linearity as a function of electric field and density, and demonstrate the enhancement of the optical non-linearity due to cooperativity.
\end{abstract}


\maketitle

Photons are robust carriers of quantum information and consequently there is considerable interest in the development of photonic quantum technologies. As optical non-linearities are extremely small at the single photon level \cite{matsuda08} attention has focussed on linear optical quantum computing \cite{knill01,kok07}. In parallel, work has been carried out on materials with a large Kerr effect \cite{hau99,mohapatra08, residori08, fushman08, schuster08} potentially enabling non-linear photonic devices. Theoretical work has explored some of the difficulties in realizing a high fidelity quantum gate based on the Kerr effect \cite{shapiro06}. An alternative mechanism for generating an optical non-linearity, for example a cooperative non-linearity due to dipolar interactions, could open new avenues for photonic quantum gates \cite{friedler05}. In a dipolar system the electric field is modified due to the local field of the neighbouring dipoles \cite{lorenztz}. Such local field effects can give rise to cooperative behaviour such as superradiance \cite{dicke54,gross82} and optical bistability \cite{hehlen94,bowden79}.

In this paper we demonstrate a cooperative optical non-linearity due to dipole-dipole interactions between Rydberg atoms. These strong interatomic interactions are sufficient to prevent excitation of neighbouring atoms to the Rydberg state \cite{saffman09} . This gives rise to a blockade mechanism which has been observed for a pair of trapped atoms \cite{isenhower09,gaetan09} and an atomic ensemble \cite{heidemann07}. In our work the effect of strong interactions between Rydberg pairs is mapped onto an optical transition using electromagnetically induced transparency (EIT) \cite{boller91,mohapatra07}. The resonant dark state responsible for EIT is modified by the dipole-dipole interactions, causing suppression of the transparency on resonance. The resulting optical non-linearity depends on interactions between pairs of atoms and is a cooperative effect where the optical response of a single atom is modified by the presence of its neighbours.

\begin{figure}
\includegraphics{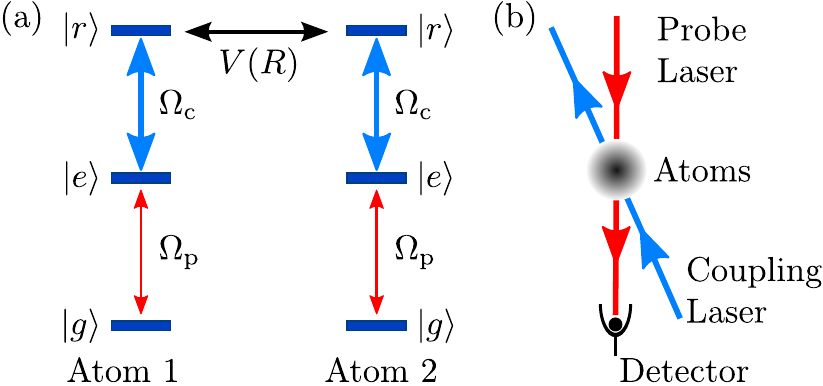}
\caption{(Color online). (a) Pair model for interacting three-level atoms. States $\vert g\rangle$, $\vert e\rangle$ and $\vert r\rangle$ are coupled by a probe field $\Omega_\mathrm{p}$ and a strong coupling field $\Omega_\mathrm{c}$. Dipole-dipole interactions couple the Rydberg states with strength $V(R)$.  (b) Schematic of experiment. EIT spectroscopy is performed on an ultracold $^{87}$Rb atom cloud.}
\label{fig:fig1}
\end{figure}

To show how dipole-dipole interactions give rise to a cooperative non-linear effect, we consider the atom pair model \cite{igor} shown in fig. 1(a) for three level atoms with ground $\vert g \rangle$, excited $\vert e \rangle$, and Rydberg $\vert r \rangle$ states. These states are coupled by a probe laser with Rabi frequency $\Omega_ \mathrm{p}$ and a strong coupling laser with Rabi frequency $\Omega_\mathrm{c}$. In the non-interacting case with probe and coupling lasers tuned to resonance the dark state is \cite{fleischhauer05}:

\begin{equation}
\vert \psi_D\rangle = \cos^2\theta \vert gg\rangle  - \sin\theta\cos\theta e^{-i\phi_r} \left(\vert gr\rangle+\vert rg\rangle \right)+ \sin^2\theta\vert rr\rangle,
\label{eq:eq1}
\end{equation}
where $\tan\theta=\Omega_\mathrm{p}/\Omega_\mathrm{c}$ and $\phi_r$ is the relative phase between probe and coupling lasers. This state is not coupled to the probe field, leading to 100~\% transparency independent of the mixing angle, $\theta$.

Dipole-dipole interactions modify this picture. The effect of interactions is to detune the doubly excited Rydberg state $\vert rr \rangle$ by the energy $V(R)$, which depends on the interatomic separation, $R$. In the limit of strong interactions ($V>\Omega_\mathrm{c}$) the detuning becomes sufficient to block excitation of $\vert rr \rangle$. This means $\vert \psi_D\rangle$ is no longer a valid eigenstate, with the new zero energy eigenstate given by \cite{moller08}:
\begin{equation}
\begin{split}
\vert \psi \rangle &= \frac{1}{\sqrt{\cos^4\theta+2\sin^4\theta}}[(\cos^2\theta - \sin^2\theta)\vert gg\rangle \\& - \sin\theta\cos\theta e^{-i\phi_r} \left(\vert gr\rangle+\vert rg\rangle \right)+ \sin^2\theta\vert ee\rangle].
\end{split}
\label{eq:eq2}
\end{equation}
This is no longer a dark state as the interaction has mixed in the rapidly decaying $\vert ee \rangle$ state in place of $\vert rr \rangle$. The result is that the EIT is suppressed by an amount which depends on the probe electric field through $\theta$. The change in the transmission, $T$, of the probe laser is due to a change in the imaginary part of the complex susceptibility at the probe laser frequency, $\chi_I=\mathrm{Im}[\chi(\omega_\mathrm{p})]$. These are related by $T=1-\exp\{-k\chi_I\ell\}$ where $k$ is the wave-vector and $\ell$ the path length through the medium. The transition between the non-interacting state eq.~(\ref{eq:eq1}) and the blockaded state eq.~(\ref{eq:eq2}) depends on the dipole-dipole coupling strength $V$. As a result, the single atom susceptibility depends on both probe electric field, giving an optical non-linearity, and also the distance to neighboring atoms, which gives a density dependent cooperative non-linearity. Hence the absorption and refractive index contributed by any atom depends on its proximity to other atoms, leading to a non-linear density scaling.

\begin{figure}
\includegraphics{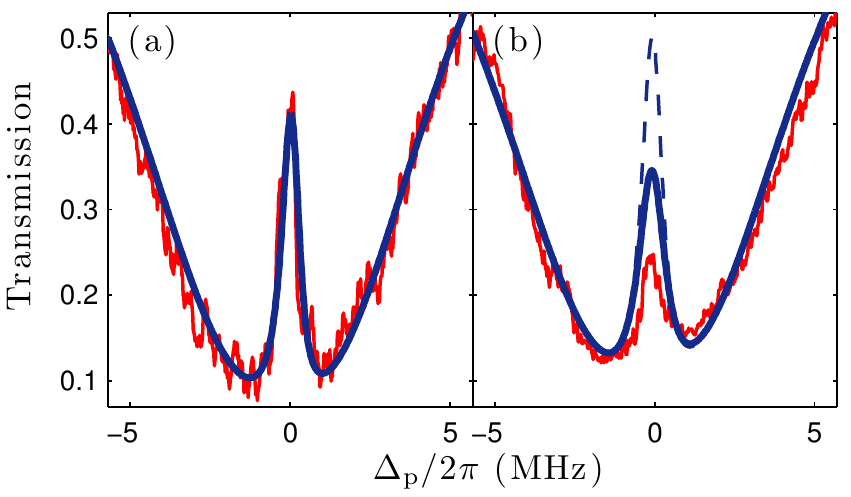}
\caption{(Color online). Suppression of EIT due to dipolar interactions for $\vert r \rangle~=~$48S$_{1/2}$ at  $N=2.2~\times10^{10}~\mathrm{cm}^{-3}$. Thin lines show absorption data at $\Omega_\mathrm{p}/2\pi$~=~0.3~MHz (a) and 1.0~MHz (b), revealing a suppression in transparency on resonance. Theory curves calculated from the pair model are plotted for $V/2\pi=10$~MHz (thick line) and $V/2\pi=0$ (dashed line) using  48S$_{1/2}$ state lifetime of $58~\mu$s \cite{beterov09}, $\Omega_\mathrm{c}/2\pi$~=~2~MHz, 300~kHz probe laser linewidth and 150~kHz relative linewidth between probe and coupling lasers.}
\label{fig:fig2}
\end{figure}

Our experiments are performed on a laser cooled $^{87}$Rb atom cloud using the experimental setup described in \cite{weatherill08} and shown schematically in fig. 1(b). Atoms are loaded into a magneto-optical trap (MOT) for 1~s after which the cooling light is extinguished. Atoms are then prepared in the 5s $^2$S$_{1/2}~\vert F=2,m_F=2\rangle$ state ($\vert g \rangle$) by  optical pumping on the 5s $^2$S$_{1/2}$ $F=2$ $\rightarrow$ 5p $^2$P$_{3/2}$ $F'=2$ transition. By varying the amount of optical pumping, the fraction of atoms in $F=2$ and hence the density in $\vert g \rangle$ can be controlled without changing the cloud size. Counter--propagating probe and coupling lasers are turned on simultaneously to perform EIT spectroscopy as shown in fig. 1(b). The probe beam absorption is detected using a photodiode with a 20~MHz bandwidth. The coupling laser has a power of $90\pm5$~mW and is frequency stabilized on resonance to the 5p $^2$P$_{3/2}~F'=3\rightarrow$ ns~$^2$S$_{1/2}$ transition using an EIT  locking scheme \cite{abel09}. The probe laser drives the 5s $^2$S$_{1/2}~F=2$ $\rightarrow$ 5p $^2$P$_{3/2}~F'=3$ transition and is scanned across the resonance from $\Delta_\mathrm{p}/2\pi = -20\rightarrow+20\rightarrow-20$~MHz in 960~$\mu$s. This double scan technique provides useful information on atom loss \cite{weatherill08}. The probe and coupling lasers are circularly polarized to drive $\sigma^+$--$\sigma^-$ transitions to maximize the transition amplitude to the Rydberg state. The path length through the ensemble along the probe axis was measured by fluorescence imaging of the cloud after preparation in $F=2$, giving $\ell = 0.52\pm0.03$~mm. Using the path length, the transmission is converted to the imaginary part of the susceptibility $\chi_I$ from the relation above. To measure the density of atoms in the probe beam, transmission data is recorded with the coupling laser off. This is then fitted using the analytic absorption profile \cite{loudon08}, giving a peak density of $N_0 = 2.2\pm0.2~\times~10^{10}~\mathrm{cm}^{-3}$. At the peak density achieved in the experiment the average pair separation is $R\simeq2~\mu$m. For states with principal quantum number $n\lesssim50$, this separation corresponds to van der Waals interactions $V(R) = -C_6/R^6$ \cite{li06}. In this regime interactions are dominated by nearest neighbours, closely approximating the pair model in fig. 1(a), unlike for resonant dipole potentials $\propto1/R^3$ for which all atoms contribute \cite{amthor09}. 

\begin{figure}[b]
\includegraphics{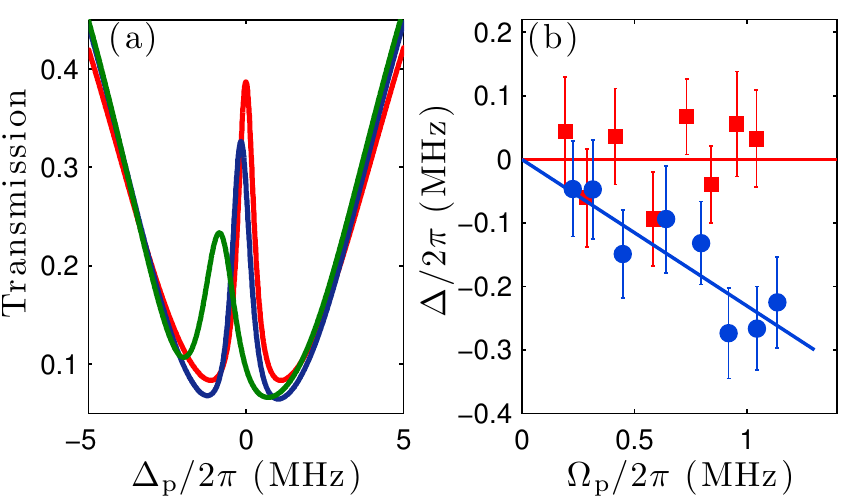}
\caption{(Color online). Effect of ions (a) Theoretical spectra for $\vert r \rangle~=~$48S$_{1/2}$ for ground state density~2.2~$\times~10^{10}$~cm$^{-3}$ for 0,~2 and 5~\% ion density showing a Stark shift in the two-photon resonance before significant suppression of the resonance. (b) Measured two-photon resonance for experimental scan (\textcolor{red}{$\blacksquare$}) shows suppression not due to ions. Reversing the probe scan direction to first create ions shows
a visible Stark shift in the second part of the scan (\textcolor{blue}{$\bullet$}), showing the sensitivity to ions. Each point is the average of 20 measurements.}
\end{figure}

Figure 2 shows individual absorption spectra recorded for $\vert r \rangle$ = 48S$_{1/2}$ at two probe powers for the first scan through resonance. This shows a significant drop in transparency as the probe Rabi frequency  is increased, demonstrating a strong suppression of the EIT resonance as expected. 

To ensure this is due to dipole-dipole interactions and not a Stark shift of the Rydberg state due to ion creation, we consider the effect of ionization on the EIT lineshape. A Monte-Carlo simulation is used to model the Stark shift experienced by atoms due to the electric field of ions within the cloud. A random distribution of atoms at a density $N_0$ is  generated and a certain fraction are randomly chosen as ions. The Stark shift for each atom is obtained by calculating the local electric field due to all ions in the cloud. The simulated EIT profile is then calculated by integrating over the Stark shifted lineshape of each atom. Figure 3(a) shows the simulated profile for $\vert r \rangle~=~$48S$_{1/2}$ calculated using a polarisability of $\alpha_0$= 38.2~MHz/(V/cm)$^2$ \cite{osullivan85}. This shows that to see a suppression of similar magnitude to fig. 2(b) due to ions, there would also be a  significant shift of the resonance ($>$1~MHz). Figure 3(b) shows experimental measurements of the two-photon resonance position as a function of probe intensity, where no shift is seen for the experimental sequence. As a further check we deliberately create ions by reversing the probe scan direction. Ions are created on the blue side of the two-photon resonance due to repulsive pair potentials \cite{amthor07a}, leading to an observable Stark shift in the second part of the scan. Since this shift is not seen for the forward probe scan, the suppression observed in fig. 2 is due to dipole-dipole interactions.

\begin{figure} [t]
\includegraphics{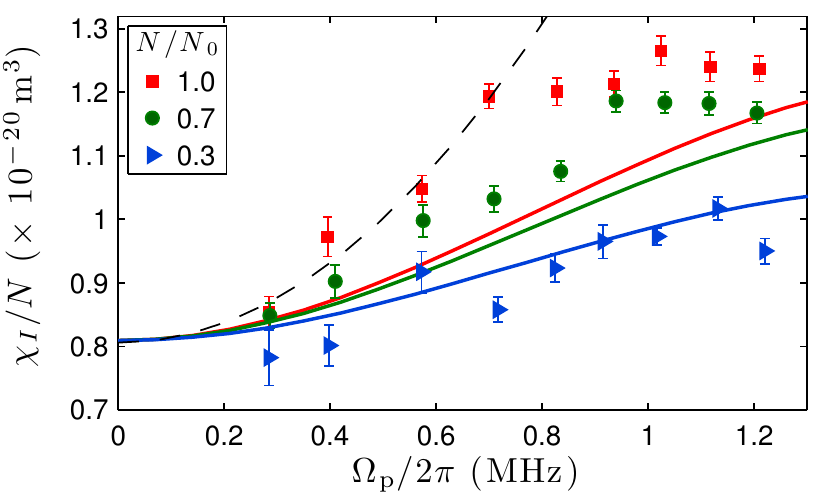}
\caption{(Color online). Optical non-linearity measured for $\vert r \rangle~=~$42S$_{1/2}$ at several densities where $N_0=2.2~\times10^{10}~ \mathrm{cm}^{-3}$, each point the average of 20 measurements. High density data is fit to a cubic non-linearity up to 
$\Omega_\mathrm{p}/2\pi$~=~0.7~MHz, shown by dashed line. This gives $\chi^{(3)}\sim2~\times~10^{-7}$~m$^2$~V$^{-2}$. Solid lines show theory curves integrating pair-model over distribution of nearest neighbours for each density, giving good qualitative agreement.}
\end{figure}

We now compare the experimental observations to theoretical predictions of the effect of dipole-dipole interactions using the pair model of fig. 1(a). To reproduce the transmission spectra shown in fig. 2, it is necessary to calculate the complex susceptibility at the probe frequency. This was done using the Liouville equation for the density matrix of the two atom system using the basis states$\{\vert gg\rangle,\vert ge\rangle,\vert eg\rangle,\vert gr\rangle,\vert rg\rangle,\vert ee\rangle,\vert er\rangle,\vert re\rangle,\vert rr \rangle \}$, where states are labelled $\vert i \rangle_1\otimes \vert j \rangle_2$. The decoherence terms included in the model are the spontaneous decay rates from the Rydberg and excited states as well as the linewidth of the probe and coupling lasers. For the Rydberg states the decay rate is a few kHz, however the relative linewidth of the probe and coupling laser is approximately 150~kHz and dominates the linewidth of the EIT feature.

Dashed lines on fig. 2(a) and (b) show theoretical absorption profiles for the non-interacting case ($V=0$), calculated using parameters optimized to fit the weak probe data at low density where interactions can be neglected. In the absence of interactions, the transparency on resonance increases with probe Rabi frequency due to population transfer from the ground state to the Rydberg state, reducing absorption. This is the opposite of what is observed experimentally, as the interactions reverse the sign of the optical non-linearity. Thick lines on fig. 2 show the absorption calculated in the blockaded regime ($V\gg\Omega_\mathrm{c}$) for $V/2\pi=10~$MHz, where increasing $V$ further gives no additional suppression. The model reproduces the enhanced absorption, however it underestimates the magnitude of the suppression. For $\vert r \rangle=48$S$_{1/2}$, the blockade radius $R_\mathrm{b}=\sqrt[6]{C_6/\Omega_\mathrm{c}}\sim4~\mu$m \cite{singer05} corresponding to an average of 5 atoms per blockade sphere. This means that we cannot only consider isolated pairs of atoms and the pair model does not give a complete description in this case. If more than two atoms contribute to the blockade then the resulting suppression increases due to a higher fraction of population in $\vert e \rangle$.

\begin{figure}[b]
\includegraphics{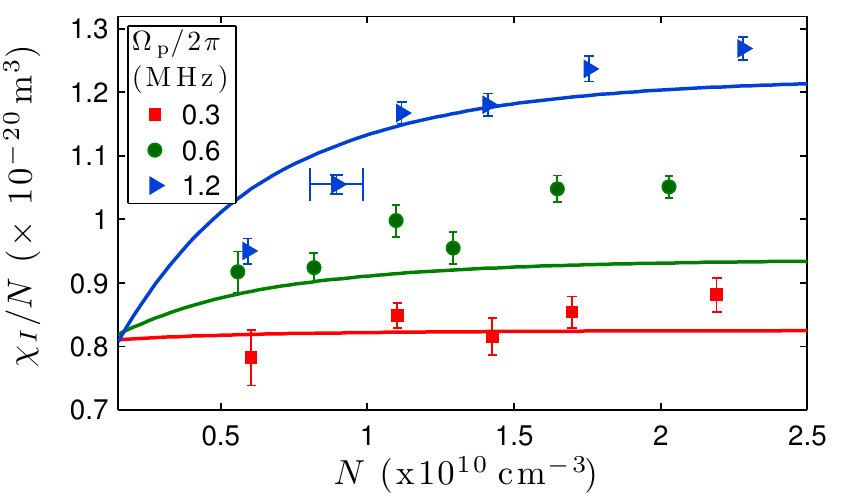}
\caption{(Color online). Cooperative non-linearity measured for $\vert r \rangle= 42$S$_{1/2}$, showing strong density dependence at high probe field. The interactions increase rapidly with $N$ due to the $1/R^6$ scaling, causing the susceptibility to plateau as $V>\Omega_\mathrm{c}$. Solid lines show model integrated over nearest neighbour distribution.}
\end{figure}

To explore the cooperative nature of the non-linearity, data were taken for $\vert r \rangle$ = 42S$_{1/2}$ which has a smaller blockade radius of 3~$\mu$m \cite{singer05}. This gives an average of 2 atoms in each blockade sphere, which should lead to better agreement with the pair model. EIT spectra are taken as a function of probe power for several densities, in each case taking 20 single measurements and extracting the average susceptibility on resonance. Figures 4 and 5 show the measured non-linearity, with the susceptibility scaled by density to give the average susceptibility of a single atom $\chi_I/N$. The optical non-linearity is shown in fig. 4, showing  that there is a rapid increase in susceptibility with probe field until it saturates at $\chi_I/N=1.2\times10^{-20}$~m$^3$ once the blockade regime is reached. To quantify the strength of the non-linearity, a third-order susceptibility is fitted to the data up to saturation, shown as a dashed line on the figure. The fit gives $\chi^{(3)}\sim2\times10^{-7}$~m$^2$~V$^{-2}$, larger in magnitude than slow light in a BEC \cite{hau99} but with two orders of magnitude lower density. This comparison illustrates the cooperative enhancement effect arising due to strong dipole-dipole interactions.

For a standard optical non-linearity the single atom response $\chi_I/N$ varies with electric field but is independent of density. In the cooperative case the response also changes with density. This is illustrated in fig.~5. For a low probe field $\Omega_\mathrm{p}/2\pi=0.3$~MHz the population remains in the ground state and is insensitive to the dipole-dipole coupling of the Rydberg state. Increasing the probe power causes the interaction effect to become important and the single atom response becomes a function of $N$.

To compare the measurements to the theoretical non-linearity predicted by the pair model it is necessary to integrate over the range of interactions strengths in the ensemble. This was done by taking the nearest neighbour distribution for a uniform density with a Poissonian atom distribution $p(R,N) = 4\pi NR^2\exp[-\textstyle{4\over3}\pi NR^3]$. For each separation $R$ the pair model was used to calculate the resonant susceptibility at a given probe Rabi frequency. We obtained the total susceptibility by summing the susceptibility weighted by $p(R,N)$ for each separation. This was repeated for each probe Rabi frequency to give the solid curves shown on figures 4 and 5, calculated using the same parameters as before except as $n$=42, $\Omega_\mathrm{c}/2\pi = 2.5$~MHz and the Rydberg lifetime is 41~$\mu$s \cite{beterov09}.

Figures 4 and 5 show the pair model agrees well with the data in the limit of low density, where interaction effects are weak, and large density or strong probe power, where the suppression reaches saturation due to blockade effects. As the saturation value of the single atom susceptibility $\chi_I/N$ agrees well with observations, this implies that modeling the system as an ensemble of isolated pairs is a valid description of the interactions, which can be used to predict the behaviour of the system. For intermediate probe powers the model underestimates the onset of suppression, seen in fig.~4, as the predicted optical non-linearity is much smaller than the measured value. This discrepancy could be explained by a modification to the nearest neighbour distribution $p(R,N)$ to consider a Gaussian rather than uniform density distribution. 

The data presented here show that the cooperativity can be controlled using both probe electric field and density. However it is also possible to modify the single atom response by changing the strength of the coupling field. As the coupling strength is increased, not only does the transparency at low density increase but stronger interactions are required to blockade excitation of Rydberg pair states. This means the density at which the cooperativity saturates can be increased.

In summary we have shown how dipole-dipole interactions between a pair of Rydberg atoms leads to a cooperative optical non-linearity. This occurs due to modification of the resonant dark state giving enhanced absorption on resonance and a modification of the dispersive properties of the medium. The cooperativity has been demonstrated through a change in the average single atom response with both density and probe electric field. On resonance, the effect of interactions is to increase the absorption of the medium, which is a  dissipative process. This could be avoided by detuning off-resonance where the interactions induce a predominately dispersive shift. In future work, this cooperative non-linearity will be used to map the blockaded atomic excitation onto the photon field, enabling the generation of non-classical states of light.

\begin{acknowledgments}
We thank M. M\"{u}ller and I. Lesanovsky for help with elucidating the theoretical origin of the suppression effect. We are also grateful to I. G. Hughes and T. Pfau for stimulating discussions, and acknowledge support from the UK EPSRC. 
\end{acknowledgments}


\begin{thebibliography}{}

\bibitem{matsuda08}
N. Matsuda \etal, Nature Photon. \textbf{292}, 95 (2008).
\bibitem{knill01}
E. Knill, R. Laflamme and G. J. Milburn, Nature \textbf{409}, 46 (2001).

\bibitem{kok07}
P. Kok \etal, Rev. Mod. Phys. \textbf{79}, 135 (2007).

\bibitem{hau99}
L. V. Hau \etal, Nature \textbf{397}, 594 (1999).

\bibitem{mohapatra08}
A. K. Mohapatra \etal, Nature Phys. \textbf{4}, 890 (2008).

\bibitem{residori08}
S. Residori, U. Bortolozzo, and J. P. Huignard, Phys. Rev. Lett. \textbf{100}, 203603, (2008).

\bibitem{schuster08}
I. Schuster \etal, Nature. Phys. \textbf{4}, 382 (2008).

\bibitem{fushman08}
I. Fushman \etal, Science \textbf{320}, 769 (2008).

\bibitem{shapiro06}
J. H. Shapiro, Phys. Rev. A \textbf{73}, 062305 (2006).

\bibitem{friedler05}
I. Friedler \etal, Phys. Rev. A \textbf{72}, 043803 (2005).

\bibitem{lorenztz}
H. A. Lorentz, Wieden. Ann. \textbf{9}, 641 (1880). L. Lorenz Wieden. Ann. \textbf{11}, 70 (1881).

\bibitem{dicke54}
R. H. Dicke, Phys. Rev. \textbf{93}, 99 (1954).

\bibitem{gross82}
M. Gross and S. Haroche, Phys. Rep. \textbf{93}, 301 (1982).

\bibitem{hehlen94}
M. P.  Hehlen \etal, Phys. Rev. Lett. \textbf{73}, 1103 (1994).

\bibitem{bowden79}
C. M. Bowden and C. C. Sung, Phys. Rev. A \textbf{19}, 2392 (1979). Y. Ben-Aryeh, C. M. Bowden and J. C. Englund, Phys. Rev. A \textbf{34}, 3917 (1986).

\bibitem{saffman09}
M. Saffman, T. G. Walker and K. M{\o}lmer, arXiv:0909.4777.

\bibitem{isenhower09}
L. Isenhower \etal, Nature Phys. \textbf{5}, 110 (2009).

\bibitem{gaetan09}
A. Ga\"{e}tan \etal, Nature Phys. \textbf{5}, 115 (2009).

\bibitem{heidemann07}
R. Heidemann \etal, Phys. Rev. Lett. \textbf{99}, 163601 (2007).

\bibitem{boller91}
K.-J. Boller, A. Imamo\u{g}lu, and S. E. Harris, Phys. Rev. Lett. \textbf{66}, 2593 (1991); A. Kasapi \etal, Phys. Rev. Lett. \textbf{74}, 2447 (1995).

\bibitem{mohapatra07}
A. K. Mohapatra, T. R. Jackson, and C. S. Adams, Phys. Rev. Lett. \textbf{98}, 113003 (2007).

\bibitem{igor}
M. M\"{u}ller and I. Lesanovsky, (private communication).

\bibitem{fleischhauer05}
M. Fleischhauer, A. Imamoglu, and J. Marangos, Rev. Mod. Phys. \textit{77}, 633 (2005).

\bibitem{moller08}
D. M{\o}ller, L. B. Madsen, and K. M{\o}lmer, Phys. Rev. Lett. \textbf{100}, 170504 (2008).

\bibitem{weatherill08}
K. J. Weatherill \etal, J. Phys. B \textbf{41}, 201002 (2008).

\bibitem{abel09}
R. P. Abel \etal, Appl. Phys. Lett. \textbf{94}, 071107 (2009).

\bibitem{loudon08}
R. Loudon, \textit{The Quantum Theory of Light} (OUP, UK 3$^\mathrm{rd}$ ed, 2008) eq.2.8.6

\bibitem{beterov09}
I. I. Beterov \etal, Phys. Rev. A \textbf{79}, 052504 (2009).

\bibitem{li06}
W. Li, P. J. Tanner, Y. Jamil and T. F. Gallagher, Eur. Phys. J. D \textbf{40}, 27 (2006).

\bibitem{amthor09}
T. Amthor \etal, Eur. Phys. J. D \textbf{53}, 329 (2009).

\bibitem{osullivan85}
M. S. O'Sullivan and B. P. Stoicheff, Phys. Rev. A \textbf{31} 2718 (1985).

\bibitem{amthor07a}
T. Amthor \etal, Phys. Rev. A \textbf{76}, 054702 (2007).

\bibitem{singer05}
K. Singer \etal. J. Phys. B. \textbf{38}, S295 (2005).

\end{thebibliography}
\end{document}